\documentclass[aps,prl,twocolumn,superscriptaddress]{revtex4-1}
\usepackage{bm}
\usepackage{amsmath,amssymb}

\usepackage{graphicx}
\usepackage{color}

\begin{document}

%\preprint{}

\title{Compact Object Mergers Driven by Gas Fallback}

\author{Hiromichi Tagawa}
\email[]{htagawa@caesar.elte.hu}
\affiliation{Institute of Physics, E{\"o}tv{\"o}s University, P{\'a}zm{\'a}ny P.s., Budapest, 1117, Hungary}
\author{Takayuki R. Saitoh}
\affiliation{Earth-Life Science Institute, Tokyo Institute of Technology, 2-12-1, Ookayama, Meguro, Tokyo, 152-8551, Japan}
\author{Bence Kocsis}
\affiliation{Institute of Physics, E{\"o}tv{\"o}s University, P{\'a}zm{\'a}ny P.s., Budapest, 1117, Hungary}

%\altaffiliation{1}{Institute of Physics, E{\"o}tv{\"o}s University, P{\'a}zm{\'a}ny P.s., Budapest, 1117, Hungary} 
%\altaffiliation{2}{Earth-Life Science Institute, Tokyo Institute of Technology, 2-12-1, Ookayama, Meguro, Tokyo, 152-8551, Japan} 

%\author{Takayuki R. Saitoh}
%\affiliation{Earth-Life Science Institute, Tokyo Institute of Technology, 2-12-1, Ookayama, Meguro, Tokyo, 152-8551, Japan}

%\author{Bence Kocsis}
%\affiliation{Institute of Physics, E{\"o}tv{\"o}s University, P{\'a}zm{\'a}ny P.s., Budapest, 1117, Hungary}

\date{\today}

\begin{abstract}
Recently several gravitational wave detections have shown evidence for compact object mergers. 
	However, the astrophysical origin of merging binaries is not well understood. 
    Stellar binaries are typically at much larger separations than what is needed for the binaries to merge due to gravitational wave emission, which leads to the so-called final AU problem. 
	In this Letter we propose a new channel for mergers of compact object binaries which solves the final AU problem. 
	We examine the binary evolution following gas expansion due to 
    a weak failed supernova explosion, neutrino mass loss, core disturbance, or envelope instability. 
	In such situations the binary is possibly hardened by ambient gas. 
	We investigate the evolution of the binary system after a shock has propagated 
	by performing smoothed particle hydrodynamics simulations. 
	We find that significant binary hardening occurs 
	when the gas mass bound to the binary exceeds that of the compact objects. 
	This mechanism represents a new possibility for the pathway to mergers for gravitational wave events. 
\end{abstract}

%\keywords{}
%fig1-0.393->275.5
%fig2-0.392->275.2
%fig3-0.768->249.3
\maketitle

\paragraph{Introduction.}
Recent gravitational wave (GW) detections show evidence for a high rate of black hole (BH)-BH and neutron star (NS)-NS mergers in the Universe
\citep*{Abbott16a,Abbott16b,Abbott17,TheLIGO17a,TheLIGO17b}. 
However, the proposed astrophysical pathways to mergers remain highly debated. 
Although several channels for the formation of close binaries have been proposed \citep[][]{Kratter11}, 
it is not obvious how separations are reduced for various types of binaries until merging due to gravitational wave emission, which leads to the so-called final AU problem \citep{Stone17}. 
The major pathways to solve this problem are 
the common envelope (CE) evolution 
which operates for 0.01-several AU separations. 
Although the binary separation can be largely reduced by the CE evolution, 
the final separations of the post CE binaries are not well understood 
\citep[][]{Ivanova13,Sabach17}. 
To reproduce the compact object mergers or x-ray binaries, 
the CE phase needs to end at some appropriate separation. 
However, many of the observed x-ray binaries 
are difficult to reproduce with the CE model; see Ref.
\citep[][]{Podsiadlowski03} and references therein. 

Here we propose a new hardening channel for mergers of compact object binaries which solves the final AU problem. 
We focus on the environments after the stellar envelope expands. 
Such a situation is expected in a weak supernova followed by 
fallback accretion, which is thought to occur 
for zero-age main sequence progenitors of masses above 
$\sim10-15~M_\odot$ \citep[][]{Woosley95,Fryer12,Sukhbold16,Raithel17}. 
Further, when the explosion fails, 
mass loss due to a neutrino emission during a protoneutron star phase 
\citep{OConnor13,Fernandez17,Coughlin17} 
causes an 
outgoing sound pulse which grows into a shock and ejects some outer material \citep{Lovegrove13}. 
Also, prior to core collapse, the hydrogen envelope may expand 
due to core disturbance \citep{Shiode14} 
or envelope instability \citep{Smith14}. 
Such precollapse expansion is observed as the luminous blue variables \citep[][]{Smith11b} or type IIn supernovae, 
which are accompanied by an expanded envelope within $\lesssim100$ AU 
and are $\sim10\%$ of all core collapse events \citep{Smith11}. 
In all these cases, 
as the gas expands and falls back, gas may exist abundantly at a large distance from the forming compact object. 

If a significant amount of gas settles around a binary, 
gas driven binary hardening occurs, either by dynamical friction (DF) 
\citep*[][]{Escala04,Chapon13,Tagawa15,Tagawa18}, 
similar to that in the CE evolution, or
via resonant disk migration \citep*[][]{Goldreich80,Kocsis11,Tang17}. 
The remarkable differences of the gas expansion and fallback scenario from the CE model are that
binaries with separations of larger than several AU can merge due to the extended gas distribution
and the merger may take place soon after this gas-driven evolutionary phase. 
In this Letter, we investigate how the binary evolves after gas expands 
and identify the range of
astrophysically important parameters required for a merger. 

\begin{figure*}
\includegraphics[width=160mm]{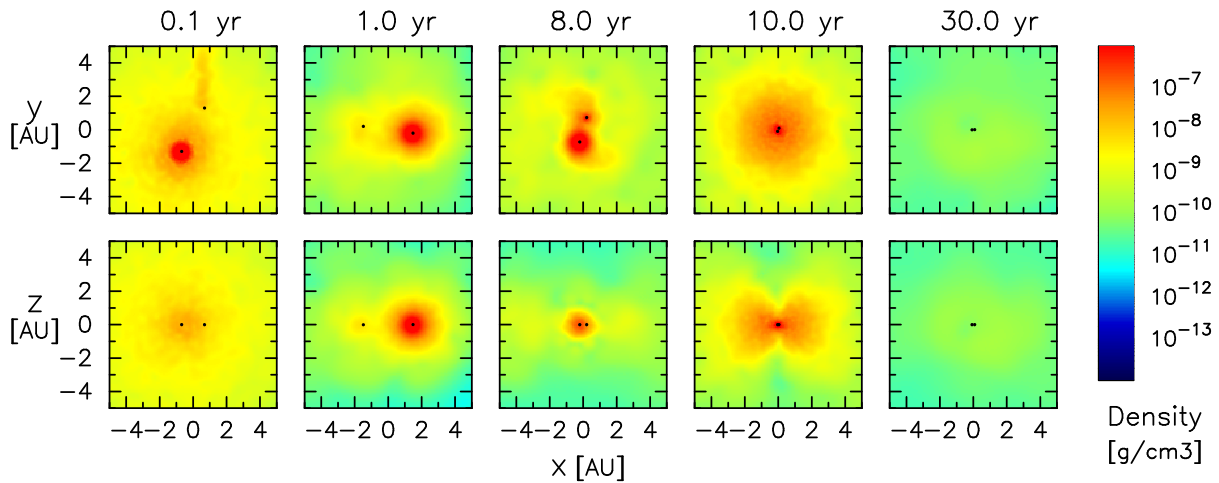}
\caption{
	The gas density maps for the fiducial model without accretion at $t=$0.1, 1, 8, 10, and 30 yr. 
    Upper and lower panels are in the $x-y$ (orbital) and $x-z$ planes, respectively. 
	Black points represent the position of compact objects. 
}
	\label{snapshot}
\end{figure*}

\paragraph{Method.}
To simulate the binary evolution interacting with ambient gas, 
we use an $N$-body/smoothed particle hydrodynamics (SPH) code, ASURA \citep*{Saitoh08,Saitoh09}. 
Gas dynamics is solved using the density-independent SPH (DISPH) method \citep*{SaitohMakino13} 
with an ideal gas equation of state. 
%Radiative cooling is not incorporated.  
We adopt a shared time step as 
the minimum of a hydrodynamical and a gravitational time step for gaseous particles 
\citep{SaitohMakino10} and a gravitational time step for the binary \citep{Baruteau11}. 

We refer to both the
explosion remnant (primary) and the companion (secondary) as compact objects (COs), 
which may be BHs, NSs, white dwarfs, or steller cores. 
Gravitational interaction among gas-gas particles is solved by a parallel tree with gravity pipe method \citep{Makino04}, 
while that between CO-gas and CO-CO is solved by a direct summation. 
The CO is modeled as a point mass which interacts only gravitationally with other particles. 
We set gravitational softening lengths 
to prevent numerical divergences during close approaches 
to $\sqrt{2}\epsilon$,  $\epsilon$, and 0 for gas-gas particles, gas-CO, 
and CO-CO interactions, respectively \citep{SaitohMakino12}.

\paragraph{Setup of simulations.}
Just after neutrino mass loss 
\citep{Fernandez17,Coughlin17} 
or precollapse expansion \citep*{Shiode14,Smith14}, or before fallback accretion \citep{Woosley95,Batta17}, 
a shock propagates to the stellar envelope and the envelope expands homologously. 
We start the simulation after a shock has propagated, 
referring the study about fallback accretion by \citet{Zampieri98}. 
Initially we set the separation between the two COs to $a_\mathrm{ini}$. 
At the beginning of the evolution, the gas cloud has homogeneous density 
within a minimum $r_\mathrm{min}=0.1$ AU 
and maximum radius $r_\mathrm{max}=0.2$ AU from the primary. 
Gas has an outgoing radial velocity profile of 
$v(r)=f_\mathrm{v} v_\mathrm{max}r/r_\mathrm{max}$ 
from the primary with 
$v_\mathrm{max}=|2\Phi_\mathrm{pri}(r_\mathrm{max})|^{0.5}
\sim220\,\mathrm{km/s}(M_\mathrm{CO,ini}/5~M_\odot)(r_\mathrm{max}/0.2\,\mathrm{AU})^{-1}$, 
where $\Phi_\mathrm{pri}(r)$ is the gravitational potential of the primary, 
$r$ is the radial distance from the primary, 
and $M_\mathrm{CO,ini}$ is the initial CO mass. 
%and $f_\mathrm{v}$ is the parameter to vary the size of an initial velocity. 

For simplicity, 
we initially rotate the binary COs and gas particles 
around the center of mass of the bound system after an initial expansion. 

In our simulations, gas accretes onto a CO 
when the gas has inward motion toward the CO, 
the gas is within $\epsilon$ from the CO, 
and the specific angular momentum of the gas relative to the CO is lower than 
the specific Keplerian angular momentum at the last stable circular orbit of a Schwarzschild BH 
as in studies of disk formation after fallback accretion \citep*[][]{Woosley12,Perna14}. 
We run additional simulations without accretion onto the COs for comparison,
which may resemble the case of strong radiation feedback, the rotation of progenitors, or precollapse expansion. 
After precollapse expansion, 
the stellar core plays the role of the primary
so the nearby gas is stabilized by gas pressure 
%exist and around a CO without accretion 
as in the standard CE evolution.
In the case without accretion, 
the gas within $\epsilon$ of the CO is not removed, although its gravitational
effects are reduced by the softening.
The simulation is stopped 
if the binary separation becomes smaller than $\epsilon$.

For the fiducial model, we set the
initial CO mass $M_\mathrm{CO,ini}=5~M_\odot$, 
the initial binary semimajor axis $a_\mathrm{ini}=3\,\mathrm{AU}$, 
the initial gas temperature $T_\mathrm{ini}=2\times10^6\,\mathrm{K}$, 
the initial gas mass $M_\mathrm{gas,ini}=10~M_\odot$, 
the gravitational softening length $\epsilon =0.1\,\mathrm{AU}$, 
the artificial viscous strength $\alpha_\mathrm{sph}=0.4$, see Eq. (46) in Ref.\cite[][]{SaitohMakino13}, 
the specific heat ratio \citep*{SaitohMakino13} of $\gamma=5/3$, 
the SPH particle number $N=3\times10^4$, 
and the neighbor number setting the kernel size of each SPH particle to be $32\pm2$. 
Both the primary and secondary CO masses $M_\mathrm{CO1}$ and $M_\mathrm{CO2}$ 
are initially set to $M_\mathrm{CO,ini}$. 
The assumed CO and gas masses correspond to zero age main sequence progenitors with mass of $15-30M_\odot$ 
whose final helium core mass is $\sim5M_\odot$ and final hydrogen mass is $\gtrsim10M_\odot$ \citep{Sukhbold16}. 
We set $T_\mathrm{ini}$ using the initial sonic velocity $c_\mathrm{s}$ ($\sim$140 km/s) at $r_\mathrm{max}=3 \times 10^{12}$ cm 
consistent with a recent study of gas dynamics following neutrino mass loss 
($\sim 100-500\,$km/s \cite{Fernandez17}). 
Furthermore, $\alpha_\mathrm{sph}=0.4$ corresponds to the viscous parameter in the $\alpha$-prescription \citep{Shakura73} of $\alpha_\mathrm{ss}\sim0.005$ \citep{Meglicki93} at $r\sim a_\mathrm{ini}$. 
We verify that the number of particles is sufficient to resolve $\epsilon=0.1\,\mathrm{AU}$ around the primary 
since the kernel size of SPH particles at $r\sim0.1$ AU is $\sim0.02$ AU. 

\begin{figure*}
\includegraphics[width=180mm]{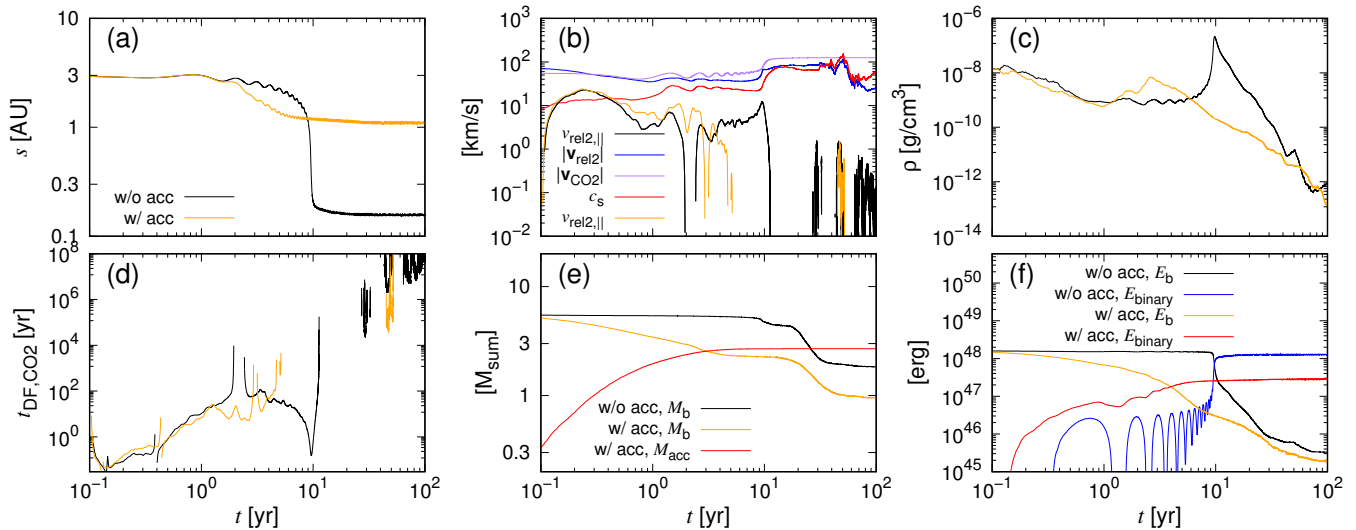}
\caption{
	Panels (a), (c), (d), and (e) show the evolution of 
	the binary separation, the gas density, 
	the dynamical friction timescale, 
	and the bound gas mass, 
    respectively, 
	for the models without (black) and with (orange) accretion. 
    Gas properties are averaged within 1-5 BHL radii of the secondary (see text).
	Blue, black, purple, and red lines in panel (b) are 
	the relative gas velocity with respect to the secondary, 
	its component
    along the secondary velocity $v_\mathrm{rel2,||}$, 
	the secondary velocity, 
	and the sonic velocity for the model without accretion, 
	the orange line shows $v_\mathrm{rel2,||}$ for the model with accretion. 
	Red line in panel (e) represents the accreted mass for the model with accretion. 
	Panel (f) shows the binding energy for the bound gas without (black) and with (orange) accretion. 
    and the binary orbital energy without (blue) and with (red) accretion. 
}
	\label{evo8_compare}
\end{figure*}

\paragraph{Dynamics of binary evolution.}
We first present the binary evolution in the fiducial model without gas accretion onto COs. 
Figure \ref{snapshot} shows the snapshots of the gas density distribution, and 
Fig. \ref{evo8_compare} shows the time evolution of the various physical quantities for the fiducial models. 
At $t=0.1$ yr of Fig. \ref{snapshot} gas just around the primary ($\lesssim$1 AU) falls back and is stabilized by gas pressure 
but the gas around the secondary is still expanding. 
Then, a gas envelope forms around the secondary 
as seen in $t=$1 yr of Fig. \ref{snapshot}. 
At this time, the Bondi-Hoyle-Lyttleton radius of the secondary 
%($GM_\mathrm{CO2}/(c_\mathrm{s}^2+v_\mathrm{CO2}^2)$) 
is $\sim$2 AU and 
the masses of envelopes mainly bound to the primary and secondary 
are $\sim5.1$ and $0.3\,M_\odot$, respectively. 
\citet{Zampieri98} indicated that there is little difference 
in the gas dynamics during fallback accretion 
between radiation-dominated completely ionized hydrogen gas   
and a polytropic gas with no radiation. 
We verify that the gas dynamics at $\sim1\,\mathrm{yr}$ 
in our simulation is 
consistent with minor differences in the density profile with respect to Ref.~\cite{Zampieri98}. 
After the gas settles, 
the binary is hardened by 
DF, i.e. the gravitational force of the density wake seen 
as tidal streamlike structures at $t=$8 yr of Fig. \ref{snapshot}. 
This causes the binary separation (black line in panel (a) of Fig. \ref{evo8_compare}) to rapidly decrease as seen in $t\sim$10 yr. 
During binary hardening, 
the gas binding energy is transferred to the binary orbital energy 
(black and blue lines in panels (f) of Fig. \ref{evo8_compare}, respectively). 
Finally, the gas expands (10 and 30 yr of Fig. \ref{snapshot}) 
and the binary hardening almost stops (black line in panel (a) Fig. \ref{evo8_compare}). 

To verify whether the main hardening mechanism is DF, 
we compare 
$t_\mathrm{hard}\sim10\,\mathrm{yr}$ to the DF timescale, 
where we define $t_\mathrm{hard}$ is %the time hardening occurs. 
the time for the binary separation 
%becomes middle between $a_\mathrm{ini}$ and the final semi-major axis $a_\mathrm{fin}$ in log scale. 
to shrink to the geometric mean of $a_\mathrm{ini}$ and $a_\mathrm{fin}$. 
Here we derive the DF timescale for the secondary, 
which has the larger velocity and angular momentum among the COs which must be
transferred to the gas for the binary to harden. 
Since DF is the gravitational force 
mostly from outside of the Bondi-Hoyle-Lyttleton (BHL) radius, 
the physical quantities, $v_\mathrm{rel2,||},~|{\bf v}_\mathrm{rel2}|,~c_\mathrm{s}$ and $\rho$, 
(see Fig. \ref{evo8_compare} for definition) 
are ad hoc averaged 
for the gas between 1 and 5 BHL radii of the secondary 
and outside of the BHL radius of the primary. 
The DF timescale for the secondary $t_\mathrm{DF,CO2}\equiv |{\bf v}_\mathrm{CO2}|/a_\mathrm{DF,CO2}$ 
(black line in panel (d) of Fig. \ref{evo8_compare}) 
is calculated using the analytic formulas of the DF acceleration derived by \citet{Ostriker99}, 
where $a_\mathrm{DF,CO2}$ is the DF acceleration for the secondary. 
We use $v_\mathrm{rel2,||}$ instead of $|{\bf v}_\mathrm{rel2}|$ in $a_\mathrm{DF,CO2}$ 
since the binary orbital decay due to DF 
is mostly caused by the force from the parallel direction to the CO motion \citep{Kim07}. 
As seen in panel (d) of Fig. \ref{evo8_compare}, 
$t_\mathrm{hard}\sim10$ yr is well matched to $t_\mathrm{DF,CO2}$ 
although $t_\mathrm{DF,CO2}$ exhibits large fluctuations due to the binary's orbital motion. 

On the other hand, in the models with accretion, $t_\mathrm{hard}$ %in the model with accretion 
is smaller than that without accretion (panel (a) of Fig. \ref{evo8_compare}). 
This is because $t_\mathrm{hard}$ is shortened by the increase of $v_\mathrm{rel2,||}$ 
due to gas accretion. 
Although $v_\mathrm{rel2,||}$ is reduced compared to $|{\bf v}_\mathrm{rel2}|$ 
by the angular momentum transfer due to the DF at $\sim0.1-2$ yr, 
this reduction is small in the model with accretion (black and orange lines in panel (b) of Fig. \ref{evo8_compare}). 
This comes from the inward motion of accreting gas in the cases with accretion. 
Since $v_\mathrm{rel2,||}$ is subsonic, the
increase in $v_\mathrm{rel2,||}$ enhances the drag force \citep{Ostriker99}. 
Thus since accretion makes DF more efficient, 
the binary hardening starts earlier 
during the accretion period of $\lesssim$5 yr 
(panels (a), (d) and (e) of Fig. \ref{evo8_compare}) in the cases with accretion. 

\begin{figure}
\includegraphics[width=90mm]{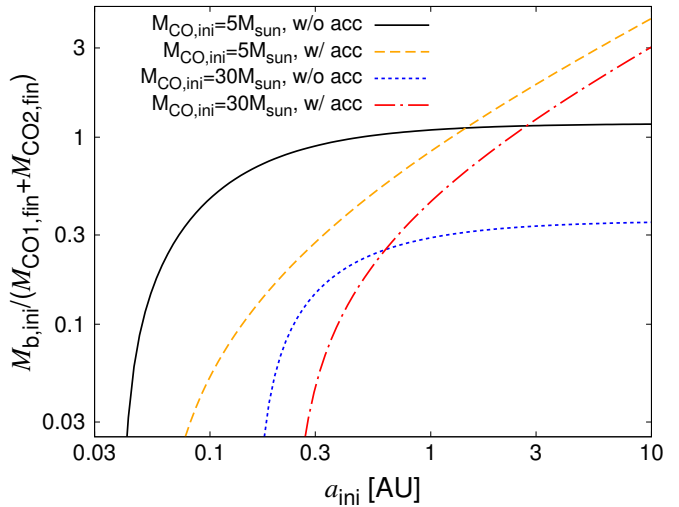}
\caption{
The criterion for the ratio of the initial bound gas mass to the initial binary mass 
above which the merger occurs within the Hubble time: 
$a_\mathrm{fin}\lesssim 
0.05\,\mathrm{AU}
(M_\mathrm{CO1}/5~M_\odot)^{1/4}
(M_\mathrm{CO2}/5~M_\odot)^{1/4}
[(M_\mathrm{CO1}+M_\mathrm{CO2})/10~M_\odot]^{1/4}$, 
assuming zero binary eccentricity for a final GW-driven merger \citep{Peters64}. 
Lines are calculated using Eq. (\ref{eq_alpha}) 
where $\alpha \lambda R_\mathrm{rlof1}$ is obtained from the simulation results, extrapolated by the derived relation between $\alpha \lambda R_\mathrm{rlof1}$ and $a_\mathrm{ini}$. 
Black and blue dashed lines show results for the fiducial models without and with accretion, respectively. 
Orange dotted and red dash-dotted lines show results for 
$M_\mathrm{CO1,ini}=M_\mathrm{CO2,ini}=30~M_\odot$ without and with accretion, respectively. 
}
	\label{mass_cri}
\end{figure}

\paragraph{Energy formalism.}
The final fate of the binary evolution may be determined as a function of physical parameters 
following the energy formalism commonly used for CE. In particular, $a_\mathrm{fin}$ is predicted as
\begin{eqnarray}
	\alpha \Bigl[ 
	G(M_\mathrm{CO1,fin}+M_\mathrm{b,fin})M_\mathrm{CO2,fin}/(2a_\mathrm{fin})\nonumber\\
-G(M_\mathrm{CO1,fin}+M_\mathrm{b,ini})M_\mathrm{CO2,fin}/(2a_\mathrm{ini}) \Bigr]=\nonumber\\
 G(M_\mathrm{CO1,fin}+M_\mathrm{b,ini})(M_\mathrm{b,ini})/(\lambda R_\mathrm{rlof1})
\label{eq_alpha}
\end{eqnarray}
\citep*[][]{Webbink84,deKool90,Ivanova13}, 
where 
$\alpha$ is the parameter that describes the efficiency of transferring the energy of the bound gas, 
$\lambda$ represents 
the typical size of the gas bound to the primary compared to $R_\mathrm{rlof1}$, 
the radius of Roche lobe around the primary \citep{Eggleton83}, 
$G$ is the gravitational constant, 
$M_\mathrm{CO1,fin}$ and $M_\mathrm{CO2,fin}$ are the final primary and secondary mass, 
$M_\mathrm{b,fin}$ is the final bound gas mass, 
$M_\mathrm{b,ini}=M_\mathrm{b,fin}+M_\mathrm{ej}$ is the initial bound gas mass, 
and $M_\mathrm{ej}$ is the ejected gas mass. 
We use $M_\mathrm{ej}$ 
after the potential energy of bound gas reaches its maximum at $\sim 1$ yr 
to reduce the contribution to Eq. (\ref{eq_alpha}) by an initial expansion. 
To investigate the dependence on $\alpha \lambda R_\mathrm{rlof1}$, we perform additional simulations for 4 models 
whose different parameters to the fiducial models with and without accretion are 
$a_\mathrm{ini}=2$~and~$5\,\mathrm{AU}$,  respectively. 
We derive fitting models, $\alpha \lambda R_\mathrm{rlof1}\propto a_\mathrm{ini}^0$ and 
$\alpha \lambda R_\mathrm{rlof1}\propto a_\mathrm{ini}^{1.2}$ for the models without and with accretion, respectively. 
In the cases with accretion, large $a_\mathrm{ini}$ increases $\alpha$ 
due to the reduction of the contribution of DF to the energy loss of gas. 
This reduction may be caused by 
the increase in accreted mass due to the decrease of the gravitational torque from the secondary,  
and the reduced DF acceleration due to
the low gas density around the binary. 
Both effects make the ratio of the energy loss by DF to that by accretion smaller.  
Using this result, we derive Fig. \ref{mass_cri} which shows the merger criterion as a function of initial parameters, Eq. (\ref{eq_alpha}) 
for different $M_\mathrm{CO,ini}$ and accretion models. 
According to Fig. \ref{mass_cri}, 
gas accretion has a large impact on hardening especially for large $a_\mathrm{ini}$. 
The final fate of the binary is determined by the bound gas mass,
which is well predicted by 
the ratio of the gravitational energy over the kinetic plus thermal energy. 
Here, in the fiducial model with and without accretion, 
$M_\mathrm{b,ini}=5.4$ and $2.7\,M_\odot$, respectively, 
$M_\mathrm{ej}=3.7$ and $1.8\,M_\odot$, 
%$M_\mathrm{acc}$=5.4 and 2.7, and 
and $M_\mathrm{b,fin}=1.7$ and $0.9\,M_\odot$. 
When the initial expansion velocity is uniformly increased by $50\%$ and $100\%$, % ($f_\mathrm{v}$=1.5 and 2), 
$M_\mathrm{b,ini}$ is, respectively, decreased by $\sim30\%$ and $\sim60\%$
in both cases with and without accretion. 
Also, when setting the initial gas density profile to be  
$\rho_\mathrm{ini} (r) \propto r^{-3}$, 
$M_\mathrm{b,ini}$ is decreased by $\sim20\%$
and increased by $\sim10\%$ 
in the cases with and without accretion, respectively, relative to that for $\rho_\mathrm{ini} (r) \propto \mathrm{const}$. 
In all of these cases, since the changes in
$\alpha \lambda R_\mathrm{rlof1}$ are modest, 
the degree of migration is predicted via Eq. (\ref{eq_alpha}) and 
Fig.~\ref{mass_cri} applies to show the condition for merger.

In the cases with gas accretion, 
the hardening rate has an upper bound 
since the ratio of the bound gas mass to the accreted mass approaches a constant as the initial gas mass increases in our settings. 
As a result, the reduction of the semimajor axis is found to be less than $\sim5$ when $M_\mathrm{CO,ini}=5~M_\odot$. 
On the other hand, in cases with negligible accretion such as for precollapse expansion, 
the hardening can be significant. 
Here the time lag between precollapse expansion and type IIn supernova 
is predicted to be $\sim1000$ yr at most \citep{Smith14}. 
By performing some additional simulations, 
in which the values of $a_\mathrm{ini}$ and $M_\mathrm{gas}$ are different from the fiducial model, 
we derive the fitting relation 
$t_\mathrm{DF,CO2}\propto \sim a_\mathrm{ini}^{3.0}M_\mathrm{b,ini}^{0.3}$ 
for $M_\mathrm{b,ini}\gtrsim M_\mathrm{b,cri}$ where $M_\mathrm{b,cri}\sim
7~M_\odot(a_\mathrm{ini}/\mathrm{AU})^{0.6}$. 
Using this relation, 
the mergers within $1000$ yr without accretion for $M_\mathrm{CO,ini}=5~M_\odot$
require $a_\mathrm{ini}\lesssim 15\,\mathrm{AU}(t_\mathrm{hard}/1000\,\mathrm{yr})^{0.33}(M_\mathrm{b,ini}/M_\mathrm{b,cri})^{-0.1}$. 
Thus there is a possibility that binaries with large separations within $\sim15$ AU 
can merge after precollapse expansion. 

%\paragraph{Neutron star mergers.}
The NS-NS merger rate inferred from the GW170817 event ($\gtrsim1\,\mathrm{yr}^{-1}$ within $100~$Mpc) 
is much larger than those estimated for 
the major merger channels such as isolated binary ($\sim6\times10^{-2}\,\mathrm{yr}^{-1}$) and dynamical evolution ($\sim2\times10^{-4}\,\mathrm{yr}^{-1}$) models \citep{Belczynski17}. 
Thus, an additional merger channel may need to be proposed to explain the observed rate 
\citep{foot1}.
%\footnote{Some estimates claim to explain the merger rate \citep{Hotokezaka17,Hotokezaka18}.}. 
Gas fallback driven mergers allow wider ranges of the initial separations for mergers. 
Natal kicks, which reduce the merger rate by about 1 order of magnitude \citep{Belczynski02} for isolated binaries, 
are small in cases of neutrino mass loss which may lead to gas fallback driven mergers. 
Thus, this channel has a potential to be highly relevant for NS-NS mergers. 
In particular, significant hardening may be caused by precollapse expansion for the remnants to remain NSs until the merger.

\paragraph{Conclusions.}
We conclude that compact object mergers may be common in cases where a significant amount of gas is released by the progenitor in the region surrounding the binary. 
The required gas mass bound to the binary must typically exceed the mass of the compact objects. 
For a $10~M_\odot$ binary, a gas-assisted merger can occur if the initial separation is $\lesssim 15$ AU, and
in cases where the COs accrete, the gas-driven reduction in the semimajor axis is $\lesssim 5$. 
We highlighted some processes that can possibly lead to these conditions, but further investigations are needed to understand the most common circumstances of this merger channel. 
Mergers facilitated by the progenitors' envelope expansion may leave an imprint on the mass and mass-ratio distribution of merger rates which will be obtained using GW detections.

\begin{acknowledgments}
We thank Scott Tremaine and Enrico Ramirez-Ruiz for useful comments. 
We are also grateful to the anonymous referees for important remarks. 
This work received funding from the European Research Council under the EU's Horizon 2020 research and innovation programme, Grant Agreement No. 638435 and 
 by 
 NKFIH KH-125675. 
Simulations and analyses were carried out on Cray XC30 and computers 
at the Center for Computational Astrophysics, National Astronomical Observatory of Japan. 
\end{acknowledgments}

{}


\begin{thebibliography}{}
\bibitem[\protect\citeauthoryear{Abbott et al.}{2016a}]{Abbott16a} 
	B. P. Abbott \textit{et al}., Phys. Rev. Lett. \textbf{116}, 061102 (2016).
    \bibitem[\protect\citeauthoryear{Abbott et al.}{2016b}]{Abbott16b} 
	B. P. Abbott \textit{et al}., Phys. Rev. Lett. \textbf{116}, 241103 (2016).
\bibitem[\protect\citeauthoryear{Abbott et al.}{2017}]{Abbott17} 
	B. P. Abbott \textit{et al}., Phys. Rev. Lett. \textbf{118}, 221101 (2017). 
\bibitem[\protect\citeauthoryear{The LIGO Scientific Collaboration et al.}{2017a}]{TheLIGO17a} 
	The LIGO Scientific Collaboration \textit{et al}., Phys. Rev. Lett. \textbf{119}, 141101 (2017).
\bibitem[\protect\citeauthoryear{The LIGO Scientific Collaboration et al.}{2017b}]{TheLIGO17b} 
	The LIGO Scientific Collaboration {\it et al}., Astrophys. J. {\bf 848}, L13 (2017).
\bibitem[\protect\citeauthoryear{Kratter}{2011}]{Kratter11} 
	K. M. Kratter, {\it Evolution of Compact Binaries. Proceedings of a workshop held at Hotel San Martin, Vina del Mar, Chile, 2011} (Astronomical Society of the Pacific, San Francisco, 2011), p. 47. 
\bibitem[\protect\citeauthoryear{Stone et al.}{2017}]{Stone17}
    N. C. Stone, B. D. Metzger, and Z. Haiman, Mon. Not. R. Astron. Soc. \textbf{464}, 946 (2017). 
\bibitem[\protect\citeauthoryear{Ivanova et al.}{2013}]{Ivanova13} 
	N. Ivanova \textit{et al}., Astron. Astrophys. Rev. \textbf{21}, 59 (2013). 
\bibitem[\protect\citeauthoryear{Sabach et al.}{2017}]{Sabach17} 
	E. Sabach, S. Hillel, R. Schreier, and N. Soker, Mon. Not. R. Astron. Soc. \textbf{472}, 4361 (2017). 
\bibitem[\protect\citeauthoryear{Podsiadlowski et al.}{2003}]{Podsiadlowski03} 
	P. Podsiadlowski, S. Rappaport, and Z. Han, Mon. Not. R. Astron. Soc. \textbf{341}, 385 (2003). 
\bibitem[\protect\citeauthoryear{Sukhbold et al.}{2016}]{Sukhbold16}
    T. Sukhbold, T. Ertl, S. E. Woosley, J. M. Brown, and  H.-T. Janka, Astrophys. J. \textbf{821}, 38 (2016). 
\bibitem[\protect\citeauthoryear{Raithel et al.}{2017}]{Raithel17} 
	C. A. Raithel, T. Sukhbold, and F. Ozel, arXiv: 1712.00021. 
\bibitem[\protect\citeauthoryear{Fryer et al.}{2012}]{Fryer12} 
	C. L. Fryer, K. Belcynski, G. Wiktorowicz, M. Dominik, V. Kalogera, and D. E. Holz, Astrophys. J. \textbf{749}, 91 (2012). 
\bibitem[\protect\citeauthoryear{Woosley \& Weaver}{1995}]{Woosley95} 
	S. E. Woosley and T. A. Weaver, Astrophys. J. Suppl. Ser. \textbf{101}, 181 (1995). 
\bibitem[\protect\citeauthoryear{O'Connor \& Ott}{2013}]{OConnor13} 
	E. O'Connor and C. D. Ott, Astrophys. J. \textbf{762}, 126 (2013). 
\bibitem[\protect\citeauthoryear{Fernandez et al.}{2017}]{Fernandez17} 
	R. Fernandez, E. Quataert, K. Kashiyama, and E. R. Coughlin, Mon. Not. R. Astron. Soc. \textbf{476}, 2366 (2018). 
\bibitem[\protect\citeauthoryear{Coughlin et al.}{2018}]{Coughlin17} 
	E. R. Coughlin, E. Quataert, R. Fernandez, D. Kasen, Mon. Not. R. Astron. Soc. \textbf{477}, 1225 (2018). 
\bibitem[\protect\citeauthoryear{Lovegrove \& Woosley}{2013}]{Lovegrove13} 
	E. Lovegrove and S. E. Woosley, Astrophys. J. \textbf{769}, 109 (2013). 
\bibitem[\protect\citeauthoryear{Shiode \& Quataert}{2014}]{Shiode14} 
	J. H. Shiode and E. Quataert, Astrophys. J. \textbf{780}, 96 (2014). 
\bibitem[\protect\citeauthoryear{Smith \& Arnett}{2014}]{Smith14} 
	N. Smith and W. D. Arnett, Astrophys. J. \textbf{785}, 82 (2014). 
\bibitem[\protect\citeauthoryear{Smith et al.}{2011b}]{Smith11b} 
	N. Smith, W. Li, J. M. Silverman, M. Ganeshalingam, and A. V. Filippenko, Mon. Not. R. Astron. Soc. \textbf{415}, 773 (2011). 
\bibitem[\protect\citeauthoryear{Smith et al.}{2011}]{Smith11} 
	N. Smith, W. Li, A. V. Filippenko, and R. Chornock, Mon. Not. R. Astron. Soc. \textbf{412}, 1522 (2011). 
\bibitem[\protect\citeauthoryear{Escala et al.}{2004}]{Escala04} 
	A. Escala, R. B. Larson, P. S. Coppi, and D. Mardones, Astrophys. J. \textbf{607}, 765 (2004). 
\bibitem[\protect\citeauthoryear{Chapon et al.}{2013}]{Chapon13} 
	D. Chapon, L. Mayer, and R. Teyssier, Mon. Not. R. Astron. Soc. \textbf{429}, 3114 (2013). 
\bibitem[\protect\citeauthoryear{Tagawa et al.}{2015}]{Tagawa15} 
	H. Tagawa, M. Umemura, N. Gouda, T. Yano, and Y. Yamai, Mon. Not. R. Astron. Soc. \textbf{451}, 2174 (2015). 
\bibitem[\protect\citeauthoryear{Tagawa \& Umemura}{2018}]{Tagawa18} 
	H. Tagawa and M. Umemura, Astrophys. J. \textbf{856}, 47 (2018). 
\bibitem[\protect\citeauthoryear{Goldreich and Tremaine}{1980}]{Goldreich80} 
	P. Goldreich and S. Tremaine, Astrophys. J. \textbf{241}, 425 (1980). 
\bibitem[\protect\citeauthoryear{Kocsis et al.}{2011}]{Kocsis11} 
    B. Kocsis, N. Yunes, and A. Loeb, Phys. Rev. D. \textbf{84}, 024032 (2011). 
\bibitem[\protect\citeauthoryear{Tang et al.}{2017}]{Tang17} 
	Y. Tang, A. MacFadyen, and Z. Haiman, Mon. Not. R. Astron. Soc. \textbf{469}, 4258 (2017). 
\bibitem[\protect\citeauthoryear{Saitoh et al.}{2008}]{Saitoh08} 
	T. R. Saitoh, H. Daisaka, E. Kokubo, J. Makino, T. Okamoto, K. Tomisaka, 
	K. Wada, and N. Yoshida, Publications of the Astronomical Society of Japan \textbf{60}, 667 (2008). 
\bibitem[\protect\citeauthoryear{Saitoh et al.}{2009}]{Saitoh09} 
	T. R. Saitoh, H. Daisaka, E. Kokubo, J. Makino, T. Okamoto, K. Tomisaka, 
	K. Wada, and N. Yoshida, Publications of the Astronomical Society of Japan \textbf{61}, 481 (2009). 
\bibitem[\protect\citeauthoryear{Saitoh \& Makino}{2013}]{SaitohMakino13} 
	T. R. Saitoh and J. Makino, Astrophys. J. \textbf{768}, 44 (2013). 
\bibitem[\protect\citeauthoryear{Saitoh \& Makino}{2010}]{SaitohMakino10} 
	T. R. Saitoh and J. Makino, Publications of the Astronomical Society of Japan \textbf{62}, 301 (2010). 
\bibitem[\protect\citeauthoryear{Baruteau et al.}{2011}]{Baruteau11} 
	C. Baruteau, J. Cuadra, and D. N. C. Lin, Astrophys. J. \textbf{726}, 28 (2011). 
\bibitem[\protect\citeauthoryear{Makino}{2004}]{Makino04} 
	J. Makino, Publications of the Astronomical Society of Japan \textbf{56}, 521 (2004). 
\bibitem[\protect\citeauthoryear{Saitoh \& Makino}{2012}]{SaitohMakino12} 
	T. R. Saitoh and J. Makino, NewA, \textbf{17}, 76 (2012). 
\bibitem[\protect\citeauthoryear{Batta et al.}{2017}]{Batta17} 
	A. Batta, E. Ramirez-Ruiz, and C. Fryer, Astrophys. J. Lett. \textbf{846}, L15 (2017). 
\bibitem[\protect\citeauthoryear{Zampieri et al.}{1998}]{Zampieri98} 
	L. Zampieri, M. Colpi, S. L. Shapiro, and I. Wasserman, Astrophys. J. \textbf{505}, 876 (1998). 
\bibitem[\protect\citeauthoryear{Perna et al.}{2014}]{Perna14} 
	R. Perna, P. Duffell, M. Cantiello, and A. I. MacFadyen, Astrophys. J. \textbf{781}, 119 (2014). 
\bibitem[\protect\citeauthoryear{Woosley \& Heger}{2012}]{Woosley12}
	S. E. Woosley and A. Heger, Astrophys. J. \textbf{752}, 32 (2012). 
\bibitem[\protect\citeauthoryear{Shakura \& Sunyae}{1973}]{Shakura73} 
	N. I. Shakura and R. A. Sunyaev, Astron. Astrophys. \textbf{24}, 337 (1973). 
\bibitem[\protect\citeauthoryear{Meglicki et al.}{1993}]{Meglicki93} 
    Z. Meglicki, D. Wickramasinghe, Bicknell G. V., Mon. Not. R. Astron. Soc. \textbf{264}, 691 (1993). 
\bibitem[\protect\citeauthoryear{Ostriker}{1999}]{Ostriker99} 
	E. C. Ostriker, Astrophys. J. \textbf{513}, 252 (1999). 
\bibitem[\protect\citeauthoryear{Kim \& Kim}{2007}]{Kim07} 
	H. Kim and W. -T. Kim, Astrophys. J. \textbf{665}, 432 (2007). 
 \bibitem[\protect\citeauthoryear{de Kool}{1990}]{deKool90} 
	M. de Kool, Astrophys. J. \textbf{358}, 189 (1990). 
\bibitem[\protect\citeauthoryear{Webbink}{1984}]{Webbink84} 
	R. F. Webbink, Astrophys. J. \textbf{277}, 355 (1984). 
\bibitem[\protect\citeauthoryear{Eggleton}{1983}]{Eggleton83} 
	P. P. Eggleton, Astrophys. J. \textbf{268}, 368 (1983). 
\bibitem[\protect\citeauthoryear{Peters}{1964}]{Peters64}
	P. C. Peters, Phys. Rev. \textbf{136}, 1224 (1964). 
\bibitem[\protect\citeauthoryear{Belczynski et al.}{2017}]{Belczynski17} 
	K. Belczynski, A. Askar, M. Arca-Sedda, M. Chruslinska, M. Donnari, M. Giersz, 
    M. Benacquista, R. Spurzem, D. Jin, G. Wiktorowicz, and D. Belloni, arXiv: 1712.00632. 
%\bibitem[\protect\citeauthoryear{Hotokezaka et al.}{2017}]{Hotokezaka17}
%K. Hotokezaka, K. Kashiyama, K. Murase, Astrophys. J. \textbf{850}, 18 (2017).
\bibitem[\protect\citeauthoryear{Belczynski et al.}{2002}]{Belczynski02} 
	K. Belczynski, V. Kalogera, and T. Bulik, Astrophys. J. \textbf{572}, 407 (2002). 
\bibitem[\protect\citeauthoryear{Hotokezaka et al.}{2017}]{Hotokezaka17}
K. Hotokezaka, T. Piran, Astrophys. J. \textbf{842}, 111 (2017).
\bibitem[\protect\citeauthoryear{Hotokezaka et al.}{2018}]{Hotokezaka18}
K. Hotokezaka, P. Beniamini, T. Piran, arXiv: 180101141
%\bibitem[\protect\citeauthoryear{Chang \& Murray}{2017}]{Chang17} 
%	P. Chang and N. Murray, Mon. Not. R. Astron. Soc. \textbf{474}, L12 (2018). 
%\bibitem[\protect\citeauthoryear{Becerra et al.}{2018}]{Becerra18} 
%	L. Becerra, C. L. Ellinger, C. L. Fryer, J. A. Rueda, and R. Ruffini, arXiv:1803.04356. 
\bibitem[\protect\citeauthoryear{}{}]{foot1} 
Some estimates claim to explain the merger rate \citep{Hotokezaka17,Hotokezaka18}.
\end{thebibliography}
\end{document}